# Venus' Mass Spectra Show Signs of Disequilibria in the Middle Clouds

Rakesh Mogul[1], Sanjay S. Limaye[2], M. J. Way[3,4,5], Jaime A. Cordova[6]

[1] Chemistry & Biochemistry Department, California State Polytechnic University, Pomona, CA, USA
[2] Space Science and Engineering Center, University of Wisconsin, Madison, WI, USA
[3] NASA Goddard Institute for Space Studies, 2880 Broadway, New York, NY, USA
[4] GSFC Sellers Exoplanet Environments Collaboration, Greenbelt, MD, USA
[5] Theoretical Astrophysics, Department of Physics and Astronomy, Uppsala University, Uppsala, Sweden
[6] Laboratory of Genetics, University of Wisconsin, Madison, WI, USA

**\*Corresponding author:** Rakesh Mogul (rmogul@cpp.edu)

**Key Points:**

- Mass data from the Pioneer Venus Large Probe Neutral Mass Spectrometer reveals several minor chemical species suggestive of disequilibria.

- Trace species in the middle clouds include phosphine, hydrogen sulfide, nitrous acid, nitric acid, hydrogen cyanide, and carbon monoxide.

- Data reveal chemicals related to anaerobic phosphorus metabolism (phosphine), anoxygenic photosynthesis (nitrite), and the nitrogen cycle.

**Abstract**

We present a re-examination of mass spectral data obtained from the Pioneer Venus Large Probe Neutral Mass Spectrometer.  Our interpretations of differing trace chemical species are suggestive of redox disequilibria in Venus' middle clouds.  Assignments to the data (at 51.3 km) include phosphine, hydrogen sulfide, nitrous acid, nitric acid, carbon monoxide, hydrochloric acid, hydrogen cyanide, ethane, and potentially ammonia, chlorous acid, and several tentative $P_xO_y$ species.  All parent ions were predicated upon assignment of corresponding fragmentation products, isotopologues, and atomic species.  The data reveal parent ions at varying oxidation states, implying the presence of reducing power in the clouds, and illuminating the potential for chemistries yet to be discovered.  When considering the hypothetical habitability of Venus' clouds, the assignments reveal a potential signature of anaerobic phosphorus metabolism (phosphine), an electron donor for anoxygenic photosynthesis (nitrite), and major constituents of the nitrogen cycle (nitrate, nitrite, ammonia, and $N_2$).

**Plain Language Summary**

We re-examined archived data obtained by the Pioneer Venus Large Probe Neutral Mass Spectrometer.  Our results reveal the presence of several minor chemical species in Venus' clouds including phosphine, hydrogen sulfide, nitrous acid (nitrite), nitric acid (nitrate), hydrogen cyanide, and possibly ammonia.  The presence of these chemicals suggest that Venus' clouds are not at equilibrium; thereby, illuminating the potential for chemistries yet to be discovered.  Further, when considering the potential habitability of Venus' clouds, our work reveals a potential signature of anaerobic phosphorus metabolism (phosphine), along with key chemical contributors towards anoxygenic photosynthesis (nitrite) and the terrestrial nitrogen cycle (nitrate, nitrite, possibly ammonia, and $N_2$).

## 1. Introduction

Venus' clouds harbor several proposed trace chemical species that suggest the potential for chemistries yet to be discovered. Exemplar trace species include ammonia, oxygen, hydrogen, methane, and ethene, which were detected remotely or in situ (*Smirnova and Kuz'min*, 1974; *Surkov*, 1977; *Oyama et al.*, 1980; *Kumar et al.*, 1981; *Moroz*, 1981; *Pollack et al.*, 1993). Recently, phosphine was reported by *Greaves et al.* (2020a), with both the detection and interpretation as a biosignature spurring significant debate within the community (*Bains et al.*, 2020; *Encrenaz et al.*, 2020; *Greaves et al.*, 2020b; *Greaves et al.*, 2020c; *Snellen et al.*, 2020; *Villanueva et al.*, 2020). In this context, we sought to examine available in situ data for signatures of trace species at Venus. Given recent interest in the potential habitability of the lower/middle cloud deck (*Limaye et al.*, 2018; *Greaves et al.*, 2020a; *Seager et al.*, 2020), we concentrated on data obtained from within the clouds by the Pioneer Venus (PV) Large Probe Neutral Mass Spectrometer (LNMS), which sampled the atmosphere during descent on December 9, 1978 (*Hoffman et al.*, 1979c).

To date, the LNMS-related literature predominantly discusses atmospheric components such as $CO_2$, $N_2$, and the noble gases with little attention given to trace/minor species, apart from methane and water (*Hoffman et al.*, 1979a; *Hoffman et al.*, 1979c; *Hoffman et al.*, 1979b; *Hoffman et al.*, 1980a; *Hoffman et al.*, 1980b; *Hoffman et al.*, 1980c; *Donahue et al.*, 1982; *Donahue and Hodges*, 1992, 1993). The LNMS data were additionally discussed by *Von Zahn and Moroz* (1985), as part of the Venus International Reference Atmosphere Model (*Kliore et al.*, 1985). A comprehensive but not exhaustive list from Venus observations (space and ground) can be found in *Johnson and de Oliveira* (2019). Beyond these studies, there is limited information on the assignment of trace chemicals, and fragmentation products, in the LNMS data.

In this study, we present a re-assessment of the LNMS mass spectral data obtained in the middle clouds (**Figure 1A**). The data in focus from an altitude of 51.3 km was originally published in identical tables in *Hoffman et al.* (1980a) and *Hoffman et al.* (1980b). In total, our interpretations match and expand upon the original LNMS studies (*Hoffman et al.*, 1979b; *Hoffman et al.*, 1980a), with these new analyses revealing the potential presence of reduced chemicals in the middle clouds including phosphine ($PH_3$), hydrogen sulfide ($H_2S$), nitrous acid ($HNO_2$), carbon monoxide (CO), ethane ($C_2H_6$), and potentially ammonia ($NH_3$), and chlorous acid ($HClO_2$). This composition is accordingly suggestive of redox disequilibria within Venus' clouds.

**2. Data and Methods**

The LNMS contained a magnetic sector-field mass analyzer (*Hoffman et al.*, 1980a), and sampled gases through a pair of metal inlet tubes (3.2 mm diameter), which were pinched at the ends that extended into the atmosphere. Data was collected from 64.2 km towards the surface, where 38 spectra were recorded at an ionization energy of 70 eV (barring the incomplete spectrum at the surface). Between ~50-25 km, the LNMS experienced a clog due to aerosol solutes, indicated as sulfuric acid by *Hoffman et al.* (1980a), which ultimately cleared at the higher temperatures at lower altitudes. The main focus of this report was spectra obtained from 64.1 to 51.3 km before the clog.

During descent, ion counts were obtained at 232 pre-selected mass positions between 1 – 208 amu and integrated over 235 ms by an on-board microprocessor. Per *Hoffman et al.* (1979c) in-flight corrections between measurements were performed using calibrants at 15 ($CH_3^+$), 68 ($^{136}Xe^{++}$), and 136 ($^{136}Xe^+$) amu to control the ion acceleration voltage and adjust for the impacts of temperature and other factors during descent. Information regarding corrections to the pre-selected amu values were not included in the archive data, nor were example *m/z* profile data, statistical insights into the measurements, or control spectra. For this study, therefore, peak shapes and shifts to the measured amu values were estimated using the LNMS count data, which contained sufficient mass points to approximate the profiles for $CH_3^+$ (15 amu), $H_2O^+$ (18 amu), $CO^+$ (28 amu), $N_2^+$ (28 amu), $^{40}Ar^+$ (40 amu), and $^{136}Xe^+$ (136 amu) – which were presumed to be pre-selected species since the respective exact masses (*Roth et* *al.*, 1976; *Haynes*, 2016) were identical or very close to the pre-selected mass values (**Supplemental Methods**).

Mass profiles from 51.3 km for these 6 pre-selected species, or references (inclusive of the $CH_3^+$ and $^{136}Xe^+$ calibrants), are displayed in **Figures 1B-F**, while those between 64.2 and 55.4 km are provided in **Figure S1**. Reasonable fits were obtained using the Gauss function (*Urban et al.*, 2014; *Stark et al.*, 2015), where regressions were unconstrained and minimized by least squares for profiles possessing >3 points per peak, and by least absolute deviations (LAD) for those with ≤3 points per peak. Regression outputs provided peak heights (calculated counts), peak means (calculated mass or amu at the centroids), and standard deviations. In turn, these terms were converted to the estimated full width half maximum (FWHM) and the difference between the calculated and expected mass (∆amu) for each respective species.

As shown in **Figure 1G**, calculated masses obtained across the altitudes of 64.2 to 51.3 km were 15.022 ± 0.001, 18.003 ± 0.001, 27.996 ± 0.001, 28.006 ± 0.003, 39.967 ± 0.004, and 135.926 ± 0.012 amu – which was indicative of the pre-selected values shifting with altitude, and increasing with increasing amu. Across 64.2-51.3 km, the total shifts (∆amu, absolute) ranged from 0.000 to 0.013 amu between 15-40 amu, and increased up to 0.030 amu at 136 amu. Across 55.4-51.3 km, the range was smaller at 0.000-0.009 amu across 15 to 136 amu. For measurements in the middle clouds (55.4-51.3 km) this was suggestive of minimal changes to the pre-selected mass positions.

Per **Figure 1H**, plotting of FWHM for the 6 references against the calculated amu revealed a linear relationship ($R^2$ = 0.994). Hence, we leveraged this trend to estimate the FWHM of poorly sampled mass peaks. For target species, the estimated FWHM and standard deviation were obtained using the linear trend at the respective altitude. In turn, regressions to poorly sampled mass peaks (<40 amu) were minimized using LAD, and constrained using the estimated FWHM (using the standard deviation as the variance) and target expected mass (using a variance that equaled the averaged ∆amu obtained between 15-40 amu at the respective altitude). Calculated FWHM values from 31 and 34 amu (at 51.3 km) are plotted as diamonds in **Figure 1H** and retain the trend of the references. Additionally, fits at 16 amu for $O^+$ and $CH_4^+$ (**Figure S2**) provided a resolving power between the mass pairs of 471 (valley minima at 12% of the O peak), which was functionally similar to the reported LNMS value of ≥440 (valley minima at 9% of O) (*Hoffman et al.*, 1980b).

Accordingly, we re-assessed the LNMS data to identify trace and minor species. In this model, chemical identities were predicated upon the assignment of atomic species, fragmentation products, and isotopologues (if so possible) to parent ions, and vice versa, where parent ions with no associated fragments, and fragments with no associated parent ions, were considered tentative. Among the many limitations to this approach, however, included a reliance upon Gauss fits and estimated FWHM, and an inability to account for potential ion scattering within the mass analyzer.

## 3. Results
### *3.1 Overview*
Parent species assigned to the LNMS data are summarized in **Table 1**; assignments are organized by the associated apparent amu, which is provided in **bold i**n the following sections for reference. Fragments and isotopologues of key parents are detailed in **Table S1**. Unique to this analysis was the identification of atomic phosphorous ($^+P$) in the data. Across the masses, isotopologues containing $^2H$ (D), $^{13}C$, $^{15}N$, $^{18}O$, $^{33}S$, $^{34}S$, and $^{37}Cl$ were observed; as were the atomic ions of $^{20}Ne$, $^{21}Ne$, $^{22}Ne$, $^{36}Ar$, $^{38}Ar$, and $^{40}Ar$. The most abundant parent ion in the data was $CO_2^+$, while polyatomic ions included $COS^+$, $SO_2^+$, and $NO_2^+$; with diatomic ions including $N_2^+$, $O_2^+$, $CO^+$, $NO^+$, $SO^+$. Assigned acidic species (weak and strong) included water, fragments consistent with sulfuric acid, and the monoprotic acids of nitrous acid ($HNO_2$), nitric acid (HNO₃), hydrochloric acid (HCl), hydrogen cyanide (HCN), and possibly hydrofluoric acid (HF), and chlorous acid (HClO₂).

Fragmentation patterns (**Table S1**) for carbon dioxide ($CO_2$) from the LNMS data and NIST mass spectral reference are displayed in **Figure S3**. The presence of $CO_2$ was evident by observation of the parent ion ($CO_2^+$), double charged parent ion ($CO_2^{++}$), all fragments ($CO^+$, $O^+$, and $C^+$), and the isotopologues of $^{13}CO_2$, $CO^{18}O$, $^{13}CO$, and $C^{18}O$. Relative abundances for $CO^+$, $O^+$, and $C^+$ were higher in the LNMS, which was suggestive of enrichment from atmospheric CO. Counts across most altitudes (barring 50-25 km, due to the clog) were supportive of a $^{13}C/^{12}C$ isotope ratio of $1.33 \times 10^{-2} \pm 0.01 \times 10^{-2}$, and $^{18}O/^{16}O$ ratio of $2.18 \times 10^{-3} \pm 0.17 \times 10^{-3}$.

**Table 1.** Assignment of parent species in the LNMS data at 51.3 km, where LNMS amu represents the pre-selected or apparent amu value.

| LNMS amu | count[a] | identity[b] | expected mass | LNMS amu | count[a] | identity[b] | expected mass |
|---|---|---|---|---|---|---|---|
| 2.016 | 22016 | **H₂** | 2.014102 | 31.990 | 327* | **O₂** | 31.990000 |
| 16.031 | 39936 | **CH₄** | 16.031300 | 33.992 | 19* | **PH₃** | 33.997382 |
| 17.026 | 244 | *NH₃* | 17.026549 | | 4* | **H₂S** | 33.987721 |
| | | *¹³CH₄* | 17.034655 | 35.005 | 6* | **PH₂D** | 35.003659 |
| 18.010 | 1200* | **H₂O** | 18.010650 | | 1* | **HDS** | 34.993998 |
| 18.034 | 20* | *NH₂D* | 18.034374 | 35.981 | 3* | **HCl** | 35.976678 |
| 20.006 | 112 | *HF* | 20.006228 | 43.991 | 1769472 | **CO₂** | 43.990000 |
| 20.015 | 30 | **H₂¹⁸O** | 20.014810 | 44.991 | 21504 | **¹³CO₂** | 44.993355 |
| | | **D₂O** | 20.023204 | 44.991 | 7936 | **CO¹⁸O** | 44.993355 |
| 27.010 | 77* | **HCN** | 27.010899 | 47.000 | 94 | **HNO₂** | 47.000899 |
| 27.988 | 423535* | **CO** | 27.995000 | 59.966 | 1 | *COS* | 59.967071 |
| 28.012 | 278529* | **N₂** | 28.012130 | 62.994 | 1 | **HNO₃** | 62.995899 |
| 28.032 | ≤50* | **C₂H₄**[d] | 28.031300 | 63.962 | 5 | **SO₂**[d] | 63.962071 |
| 28.997 | 7040* | **¹³CO** | 28.998355 | 65.961 | 0.3* | **³⁴SO₂**[d] | 65.957867 |
| 29.997 | 940* | **C¹⁸O** | 29.999160 | 67.964 | 6272 | *HClO₂* | 67.966678 |
| 30.046 | ≤100* | **C₂H₆** | 30.046950 | 78.053 | 7* | **C₆H₆** | 78.046950 |
| 31.972 | ≤8* | **³²S** | 31.972071 | 80.947 | 1 | *NSCl* | 80.943998 |

**(a)** observed and calculated counts
**(b)** italics: tentative assignment
**(d)** parent and/or fragment ion
* calculated counts

### *3.2 Hydrogen Sulfide & Phosphine*

**Table S1** lists the mass data (51.3 km) and assignments for hydrogen sulfide ($H_2S$) and/or phosphine ($PH_3$) along with the associated fragments and isotopologues. Due to similar masses for $H_2S^+$ and $^+PH_3$, as well as $HS^+$ and $^+PH_2$ ($\Delta m$ values of 0.009661), resolving powers beyond the capabilities of the LNMS (3519 and 3414) would be required for separation. Therefore, unambiguous assignments for the parent ion ($M^+$) and first fragmentation product ($[M-H]^+$) were not possible. Rather, identities were assigned using the following rationale:

1. Assignment of ≤10 counts for $S^+$ at 51.3 km was predicated upon devolving isobaric $O_2^+$, which was the dominant species at the mass pair of **32.972** and **32.990 amu**. Per **Figure 1I**, regressions (LAD) using the described constraints provided 0 counts for $S^+$, which implied an absence of $H_2S^+$; however, fits to the data were modest, per the summed absolute deviation (SAD) value of ~59 (where SAD values approaching zero indicated better fits). Per **Figure 1J**, fits were better minimized (SAD, ~0.04) when using expanded constraints (averaged $\Delta$amu plus the standard deviation, and 2x the standard deviation of the estimated FWHM), which provided 10 counts.

2. Counts were assigned to $^+P$. Per **Figures 1K-N**, fits to the mass pair at **30.973** & **31.006 amu** revealed discernable peaks with differing peak ratios for $P^+$ and $HNO^+$ across the altitudes of 59.9-51.3 km. Regressions (LAD) to the mass pair across these altitudes were maximally minimized when including two species, $P^+$ and $HNO^+$, as indicated by the range in SAD values of $1.6 \times 10^{-6} - 9.8 \times 10^{-5}$. In comparison, fits using only $HNO^+$ provided SAD values of ~2.1 – 5.7. Calculated counts for $P^+$ were the highest at 51.3 km, and below the detection limit at ≥61.9 km. This suggested the presence of a heterogeneously-mixed, phosphorus-bearing, and neutrally-charged parent gas or vapor in the middle clouds.

3. Across 64.1-50.3 km, the counts at 34 amu represented $H_2S^+$ (33.987721 amu), $^+PH_3$ (33.997382 amu), or a composite of $H_2S^+$ and $^+PH_3$. Reasonable fits were obtained across the mass triplet at 34 amu (**33.966**, **33.992**, & **34.005 amu**) when using single species (**Figures 1O-P**). For H$_2$S$^+$ or $^+$PH$_3$ (at 51.3 km), SAD values from the regressions (LAD) were both ~4.7. Surprisingly, inclusion of a composite provided a better fit and lower relative SAD value of ~4.0 (at 51.3 km), per **Figure 1Q**, with regressions (LAD) yielding 18% H$_2$S (4 counts) and 82% PH$_3$ (19 counts).

4. The counts of 18 at **32.985 amu** represented HS$^+$ (32.979896 amu), $^+$PH$_2$ (32.989557 amu), or a composite of HS$^+$ and $^+$PH$_2$.

5. The fragment $^+$PH (31.981732 amu) was not available for detection due to being masked by $^+$O$_2$ (**31.990 amu**; counts = 356), which was ~30-fold higher in abundance than the $^+$PH$_3$ parent ion (calculated counts = 10), and >200-fold higher than the expected counts of ~1.5 for $^+$PH, per the NIST reference.

6. When considering deuterium and other isotopologues (51.3 km), the counts of 12 at **35.005 amu** were attributed to HDS$^+$ (33.987721 amu), $^+$PH$_2$D (33.997382 amu), and/or H$_2$$^{33}$S (34.987109 amu).

7. In the data, when considering the conditions of the middle clouds (~74 °C, ~ 1 bar, ~50 km), no other parent neutral gases, other than PH$_3$, could fully account for the presence of $^+$P. Alternative gaseous/vaporous and mineralized chemicals included PCl$_3$, H$_3$PO$_4$, and P$_2$O$_5$; however, these species could not be fully accounted for in the data, or were considered incompatible with the LNMS inlets:

    a. For PCl$_3$, (1) the parent ion of $^+$PCl$_3$ was isobaric with $^{136}$Xe$^+$ (a high abundance calibrant) and could not be confirmed, (2) the mass for $^+$PCl$_2$ (100.911612 amu) likely corresponded to 0 counts (or counts of <0.5) at **100.990 amu** (51.3 km), and (3) the mass for $^+$PCl (65.942760 amu) was not sampled by the LNMS.

    b. For H$_3$PO$_4$, many of the potential fragment ions were isobaric with SO$_x$$^+$ species and/or possessed counts of 0 (at 51.3 km): (1) at **47.966 amu**, the counts of 10 were

attributed to SO$^+$ (47.967071 amu) and/or HPO$^+$ (47.976732 amu), (2) at **63.962 amu**, counts of 5 were attributed to SO$_2^+$ (63.962071 amu) and/or HPO$_2^+$ (63.971732 amu), (3) at **79.958 amu**, counts of 0 (or <0.5) were attributed to SO$_3^+$ (79.957071 amu) and/or HPO$_3^+$ (79.966732 amu), and (4) 0 counts were recorded at masses (**81.975** & **96.667 amu**) potentially attributed to H$_3$PO$_3^+$ and H$_2$PO$_4^+$ (81.982382 & 96.969557 amu).

c. For H$_2$SO$_4$, which is structurally similar to H$_3$PO$_4$, fragmentation to S$^+$ occurs in insignificant yields (≤1% of the parent ion; <0.5% of the base peak, SO$_3^+$) per the NIST reference. Similarly, initial calculations suggest a ~6% yield for S$^+$ from H$_2$SO$_4$ during the clog (36.8 km), where the LNMS was postulated to be enriched in sulfuric acid fragments. By extension, yields for fragmentation of H$_3$PO$_4$ to P$^+$ may also be very low.

d. For P$_2$O$_5$, while mass profile likely overlapped with **142.486 amu** (2 counts), survival of the parent ion through the pinched and low-conductance gas inlets was considered to be unlikely (similar to H$_2$SO$_4$).

In summary, for PH$_3$, measured counts at 51.3 km potentially correlated to $^+$P, $^+$PH$_2$, $^+$PH$_3$, and $^+$PH$_2$D; while counts for $^+$PH were masked by O$_2$. In this analysis, no other viable parent ions could account for $^+$P – though this did not exclude $^+$P arising from a dissociated H$_x$P$_y$O$_z$ species. For H$_2$S, measured masses correlated to S$^+$, HS$^+$, H$_2$S$^+$, and HDS$^+$, with regressions indicating 0-10 counts for S$^+$, and minimal abundances of $^{34}$S$^+$. Regressions to the mass triplet at 34 amu (**Figures Q-S**) were supportive of the following composites listed in order of increasing altitude:

- At 50.3 km (SAD, ~1.1), where the clog began to occur during descent: ~50% H$_2$S$^+$ and ~50% PH$_3^+$ (~4 counts each).
- At 51.3 km (SAD, ~4.0): ~18% H$_2$S$^+$ (~4 counts) and ~82% $^+$PH$_3$ (~19 counts).

- At 55.4 km, through relatively moderate fits (SAD, ~4.8): ~28% $H_2S^+$ (~2 counts) and 72% $^+PH_3$ (~5 counts).
- Between 58.3-61.9 km (SAD <3.7): A decrease from ~4 to ~2 for $H_2S^+$, and a range of ~0-0.1 for $^+PH_3$.
- At 64.2 km: Both ions were below or at the detection limit.

Comparison to the NIST spectral references revealed similar fragmentation patterns for $^+PH_3$ and $H_2S^+$ (**Figures 1T-U**), respectively, where the higher yields for $HS^+$ and $^+PH_2$ were supportive of counts arising from fragmentation of $HDS^+$ and $^+PH_2D$. Relative abundances for $^+P$ were similar to the NIST reference, while expected abundances of $S^+$ (~2 counts, 51.3 km) fell within the range of 0-10 counts from regressions. Across the regressions, the counts obtained at 31, 32, and 34 amu were complementary and supportive of the model.

For the deuterium isotopologues, across 59.9 to 50.3 km, substantially high D/H ratios with large propagated errors (~50-400%) were obtained due to the low counts. For these calculations, counts for $PH_2D$ and $HDS$ at each altitude were corrected for $^{35}Cl^+$ (using fits across the mass pair at 35 amu), corrected for $H_2^{33}S^+$ (using the $^{33}S/^{32}S$ ratio obtained from $^{33}SO^+$ and $^{32}SO^+$; **Section 3.4**), and disambiguated using the calculated $PH_3^+/H_2S^+$ ratio given the similar degrees of hydrogen-deuterium exchange for $H_2S$ and $PH_3$ (*Jones and Sherman*, 1937; *Weston Jr. and Bigeleisen*, 1952; *Wada and Kiser*, 1964; *Fernández-Sánchez and Murphy*, 1992). At 51.3 km, this provided high ratios of $1.6 \times 10^{-2} \pm 1.0 \times 10^{-2}$ for $HDS/H_2S$ and $1.0 \times 10^{-2} \pm 0.5 \times 10^{-2}$ for $PH_2D/PH_3$, which were suggestive of underestimations of the $H_2^{33}S$ abundances and/or variance in the lower counts. In support, the composite isotopologue ratio (($PH_2D+HDS$)/ ($PH_3+H_2S$)) at 51.3 km was 0.63 and decreased >1000-fold to 0.045 ± 0.021 between ~24 to 0.9 km, where counts were substantially larger, and representative of less statistical variation. This composite isotopologue ratio was calculated using uncorrected counts at **35.005 amu** ($PH_2D+HDS$) and **33.992** (PH$_3$+H$_2$S). We note that *Donahue and Hodges* (1993) reported a similar ratio of 0.05 for HDS/H$_2$S using the same mass points and uncorrected counts below the clouds.

*3.3 Brønsted-Lowry Acids*

The LNMS data contained counts for masses consistent with HNO$_3^+$, HNO$_2^+$, NO$_2^+$, HNO$^+$, NO$^+$, $^+$OH, O$^+$, and N$^+$ (**Table S1**). Devolved plots at 31 amu (**Section 3.2**) supported the presence of HNO$^+$. Fragmentation products of HNO$_3$, per published reports (*Friedel et al.*, 1959; *O'Connor et al.*, 1997), do not include HNO$^+$ or HNO$_2^+$. For HNO$_2$, we found conflicting evidence for HNO$^+$ as a fragmentation product; with spectra from PubChem (CID 386662) supporting HNO$^+$ (**Figure S4**). Together, this was suggestive of HNO$_3^+$ and HNO$_2^+$ being parent ions, and nitroxyl hydride (HNO$^+$) being a fragment of HNO$_2^+$. Counts for NO$^+$ (~413) and NO$_2^+$ (≤620), the base peaks of HNO$_2^+$ and HNO$_3^+$, were estimated by disambiguating the isobaric species of C$^{18}$O$^+$ and CO$^{18}$O$^+$, respectively (**Supplemental Methods**). Per **Figure S4**, fragmentation patterns for HNO$_2^+$ followed the general trend of the reference. Potentially co-present and isobaric species included PO$^+$ (46.968910 amu) and PO$_2^+$ (62.963907 amu).

The mass data also revealed assignments for HCl and HCN (**Table S1**), and possibly HF and HClO$_2$. Across the mass pair of **35.966** and **35.981 amu**, H$^{35}$Cl was a potential minor component against the major isobar of $^{36}$Ar$^+$. Similarly, at **37.968 amu**, H$^{37}$Cl was a minor component against the major isobar of $^{38}$Ar$^+$. At 51.3 km, fits to 35 amu (**34.972 & 35.005 amu**) provided calculated counts of 12 for $^{35}$Cl$^+$. When assuming HCl to be the parent source, per yields from the NIST reference, this amounted to ~80 counts for H$^{35}$Cl$^+$. For $^{37}$Cl$^+$, counts were corrected for C$_3$H$_1^+$, a benzene fragment, (see **Supplemental Methods**), and for D$^{35}$Cl$^+$ using a (D/H)$_{HCl}$ ratio of 0.0303 from 74 km (*Krasnopolsky et al.*, 2013). Across the altitudes of 58.3-51.3 km, this yielded a $^{37}$Cl/$^{35}$Cl ratio of 4.5x10$^{-1}$ ± 0.7x10$^{-1}$, comparable to the terrestrial value (**Table S2**).

At the respective positions of 26 and 27 amu, HCN$^+$ (27.010899 amu) and CN$^+$ (26.003074 amu) were likely the dominant species, with HCN being the dominant parent source.  At best, the isobaric species of $C_2H_3^+$ and $C_2H_2^+$ (see **Section 3.6**) were minor constituents in peak profiles for $HCN^+$ and $CN^+$, respectively.  Similarly, $HF^+$ was likely a minor component against the isobaric species $H_2{}^{18}O^+$ and $^{20}Ne^+$; as was $F^+$ against the isobaric species of $^{18}OH^+$ and $^{40}Ar^{++}$.  For $HClO_2$, the counts at **50.969** amu was consistent with assignment of $ClO^+$ (50.963853 amu), while counts at **66.963** and **67.964 amu** were tentatively assigned to composites of $ClO_2^+$ (66.958853 amu) with $^{134}Xe^{++}$ (66.952697 amu), and $HClO_2^+$ (67.966678 amu) with $^{136}Xe^{++}$ (67.953610 amu), respectively.

*3.4 Oxysulfur Species*

The data supported the presence of several potential fragments of $H_2SO_4$ (with counts ranging from 2-10) including $^{33}SO^+$, $^{34}SO^+$, $^{33}SO_2^+$, $^{34}SO_2^+$, and potentially $HSO_2^+$ (**Table S1**). Fragmentation yields (**Figure S4**), however, were dramatically different from the NIST reference (no counts were observed for $H_2SO_4^+$) due to the impact of viscous flow through the crimped inlets of the LNMS, which promoted dissociation of $H_2SO_4$ at the inlet prior to entering the ion source (*Hoffman et al.*, 1980a).  Given the presence of water in the data, the $H_2SO_4$ was presumably acquired from aerosolized species and not vapor.  Per **Section 3.1** (Step 6b), $SO_3^+$, $SO_2^+$, and $SO^+$ were respectively isobaric to $HPO_3^+$, $HPO_2^+$, and $HPO^+$.  However, counts from $SO^+$ were suggestive of $^{33}S/^{32}S$ and $^{34}S/^{32}S$ ratios of $1.3 \times 10^{-2} \pm 0.9 \times 10^{-2}$ and $5.9 \times 10^{-2} \pm 0.8 \times 10^{-2}$ across the altitudes of 39.3 and 24.4 km, respectively, where the LNMS was postulated to be enriched in sulfuric acid fragments.  For $SO_2^+$, this amounted to disentangled counts of 0.3 for $^{34}SO_2^+$, which indicated that $^{132}Xe^{++}$ (65.952077 amu) was the major species at **65.961 amu**. Similarly, the isobars of $^{130}Xe^{++}$, $^{33}SO_2^+$, and $HSO_2^+$ were likely mixed at **64.960 amu**.  The LNMS data suggested the presence of several Xe isotopes including $^{128}Xe$, $^{129}Xe$, $^{130}Xe$, $^{132}Xe$, and $^{134}Xe$.

*3.5 Ammonia*

**Table S1** lists the mass data and assignments potentially consistent with $^+NH_2D$, $^+NH_3$ (ammonia), $^+NH_2$, $^+NH$, $^+N$, and the isobars of water and methane-related species.  For water (**Figures 1C & S1**), fits to the mass triplet at 18 amu (**17.985**, **18.010**, & **18.034 amu**) were best minimized when including ~20 counts from $^+NH_2D$ (18.032826 amu); inclusion of $^+NH_2D$ yielded no changes to the FWHM for water or averaged $\Delta$amu (at 51.3 km).  The assignment of $^+NH_2D$, while tentative, was supportive of $NH_3^+$ being the parent species. In the data, the mass at **17.026 amu** was consistent with $^+NH_3$ (17.026549 amu); however, $^{13}CH_4^+$ (17.034655 amu) was the dominant species at this position (due to use of $CH_4$ as a calibrant). Likewise, masses potentially consistent with $^+NH_2$ (**16.019 amu**) and $^+NH$ (**15.013 amu**) were dominated by the isobars of $^{12}CH_4^+$ (16.031300 amu) and $^{12}CH_3^+$ (15.023475 amu), respectively.

### 3.6 Low-Mass Organics

**Table S1** lists the mass data and assignments for methane ($CH_4$), ethane ($C_2H_6$), benzene ($C_6H_6$), and related fragments. Across all altitudes (64.2-0.9 km), fragmentation of $CH_4$ yielded $CH_3^+$ in relative abundances of 76 ± 6%, similar to the NIST and MassBank references (~83-89%). Atomic carbon in the LNMS data was enriched as expected due to tremendous input from $CO_2$.

For ethane ($C_2H_6$), the LNMS data possessed pre-selected masses consistent with $C_2H_6^+$, $C_2H_5^+$, $C_2H_4^+$, $C_2H_3^+$, and $C_2H_2^+$. Per our model, these species were likely minor components against the isobaric alternatives of $C^{18}O^+$, $^{13}CO^+$, $N_2^+$ and $CO^+$, $CN^+$, and $HCN^+$, respectively. Fits to 30, 27, and 26 amu were indicative of $C_2H_6^+$ (≤100 counts), $C_2H_3^+$ (≤50 counts) and $C_2H_2^+$ (≤10 counts) being constituents of the peaks dominated by $^{12}C^{18}O^+$, $HCN^+$, and $CN^+$, respectively.

For benzene, the data (at 51.3 km) possessed counts corresponding to $C_5^{13}CH_6^+$ (($M+1)^+$), $C_6H_6^+$ ($M^+$), and $C_6H_5^+$ ($[M-H]^+$) at **78.924**, **78.053**, and **77.040 amu**, respectively. As described in the **Supplemental Methods**, disambiguation of the counts suggested the presence of isobaric species at **78.053 amu** such as dimethyl sulfoxide (($CH_3)_2SO^+$) and/or $P_2O^+$. Lastly, in this model, $C_3H_4$ (propyne) was indistinguishable against the mass peak for $^{40}Ar^+$ (**Figure 1E**).

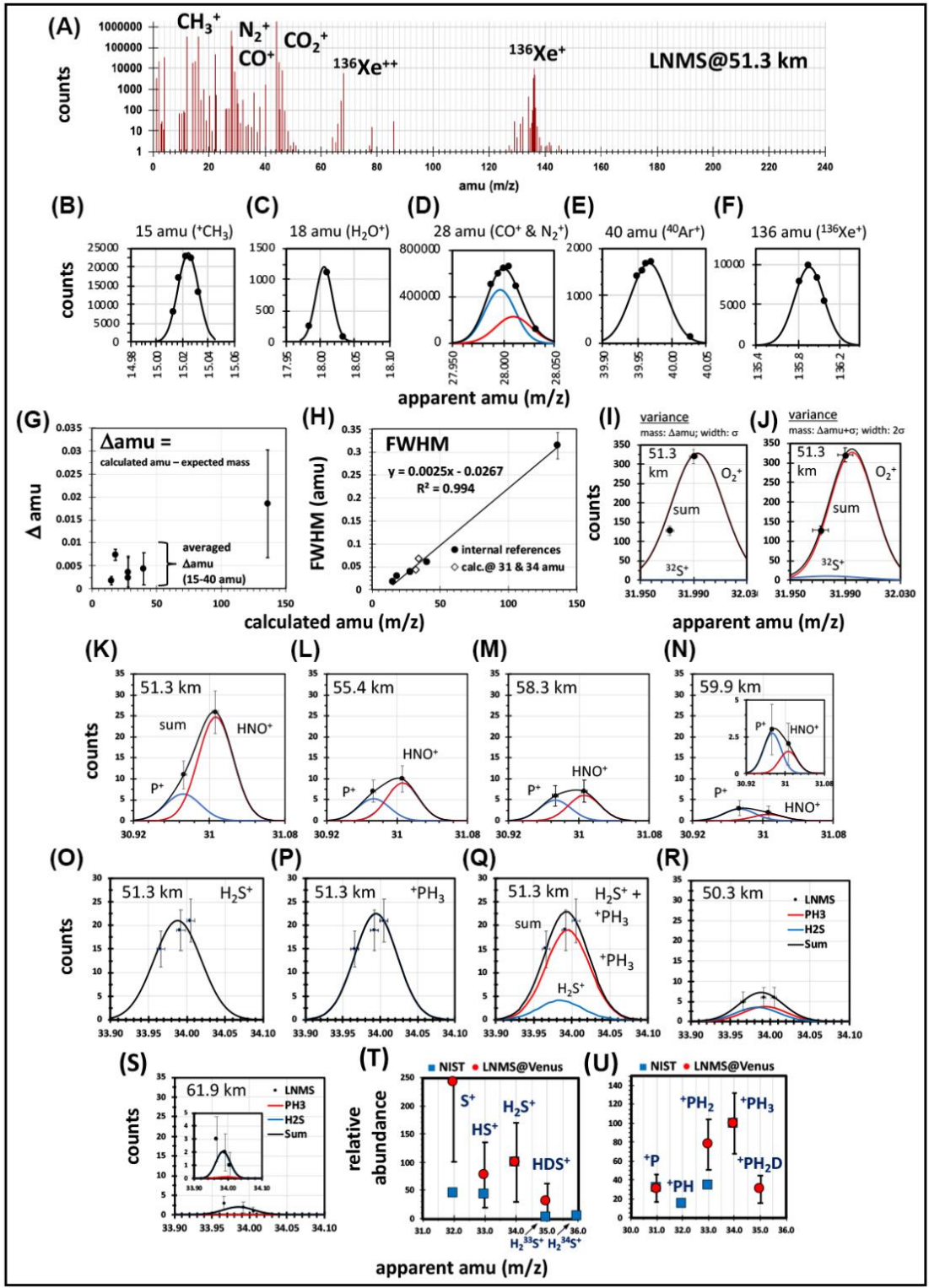

*Figure 1.* (A) LNMS spectra obtained at 51.3 km with annotations for the major species and in-flight calibrants. (B-F) Approximate peak shapes at 51.3 km obtained from regressions of the mass points at 15 amu ($CH_3^+$), 18 amu ($H_2O^+$), 28 amu ($CO^+$ & $N_2^+$), 40 amu ($^{40}Ar^+$), and 136 amu ($^{136}Xe^+$); y-axis error bars are smaller than marker size of the data points. (G-H) Relationships between calculated amu and Δamu (Δamu = calculated amu – expected mass) and full width half maximum (FWHM), where averages and standard deviations (error bars) were calculated across the altitudes between 64.2 and 51.3 km (most error bars are smaller than the marker size); diamonds represent the calculated FWHM from deconvolutions at 31 and 34 amu. (I-J) Fits to the mass pair at 32 amu at 51.3 km showing $^{32}S^+$ (blue), $O_2^+$ (red), and summed value (black) using differing variances for the mass and FWHM terms. (K-N) Fits to the mass pair at 31 amu for $P^+$ (blue), $HNO^+$ (red), and summed value (black) across 59.9 to 51.3 km; x-axis error bars represent the standard deviation for the averaged Δamu (between 15-40 amu) at the respective altitude, and y-axis error bars represent the square root of the counts. (O-Q) Fits to the mass triplet at 34 amu from 51.3 km for $PH_3$ (red), $H_2S$ (blue), and a composite of $PH_3$ and $H_2S$; plot layout and error bars are as described above. (R-S) Fits to the mass triplet at 34 amu from 50.3 and 55.4 km for a composite of $PH_3$ (red )and $H_2S$ (blue); plot layout and error bars are as described. (T-U) Comparison of fragmentation patterns for $PH_3$ and $H_2S$ from the LNMS data (red circles) and the respective NIST mass spectral references (blue squares); counts for $S^+$, $H_2S^+$, $HDS^+$, $^+P$, $^+PH_3$, and $^+PH_2D$ were obtained as described, while counts for $^+PH_2$ and $HS^+$ were disambiguated using the relative abundances of the parent species; error bars represent the square root of the counts, and masses are displayed in unit resolution for clarity.

## 4. Discussion

Assignments to the LNMS data reveal fragmentation products and parent ions that support the presence of novel chemical species in Venus' atmosphere. Atomic phosphorus was among the assignments; thereby, indicating the presence of a phosphorus-bearing gas or vapor in Venus' clouds. Across the altitudes of 58.3-51.3 km, phosphine ($PH_3$) represented the simplest phosphorus-bearing gas that fit the LNMS data best. While $H_3PO_4$ remains a viable candidate, matches to the combined data require very high vaporous or aqueous aerosol abundances relative to $H_2SO_4$. Additionally, while counts in the data support the presence of $P_2O_5$, a proposed suspended mineral in Venus' clouds (*Krasnopolsky*, 1989), it is our understanding that the LNMS inlets were designed to restrict entry of such types of molecules. We were also unable to find literature precedent for $P_2O_5^+$ as a parent ion under conditions similar to the LNMS or NIST references. Alternative gaseous candidates included (1) phosphorus trichloride ($PCl_3$), which was inadequately described by the data, (2) elemental phosphorus ($P_4$), phosphorus dioxide ($PO_2$), and phosphorus monoxide ($PO$), which are not gases under the conditions of Venus' clouds, and (3) diphosphorus oxide ($P_2O$), which is an unstable gas that potentially serves as dissociative (*e.g.*, at the inlet) and/or fragmentation product from a larger $P_xO_y$ species; however, we were unable to find literature precedent for $P_2O^+$ as a mass spectral fragmentation product or parent ion. Thus, we propose that phosphine and $H_2S$ are potentially co-present in the middle clouds.

The LNMS data also support the presence of acidic species including $HNO_2$, $HNO_3$, HCl, HCN, and possibly HF and $HClO_2$. The presence of $HNO_2^+$ is supported by assignments of the fragment products of $HNO^+$ and $NO^+$, and preliminary analyses show that $HNO_2^+$ and $HNO^+$ track well across the altitude profile towards the surface. Assignments of $NO_2^+$ and $NO^+$ support $HNO_3^+$, where counts for $HNO_3^+$ substantially increase from 1 (at 51.3 km) to ~720 below the clouds. When considering all potential nitrogen parent species, the LNMS data support a range in nitrogen oxidation numbers (or states) including -3 (HCN and possibly $NH_3$), 0 ($N_2$), +3 ($HNO_2$), and +5 ($HNO_3$).

The LNMS data additionally show the presence of CO, $O_2$, and possibly COS and NSCl. While parent ions of COS and NSCl were observed, no fragmentation products could be identified, other than the atomic ions. Comparison of the $CO_2$ fragmentation patterns revealed a $CO^+/CO_2^+$ ratio of ~0.16-0.24 at 51.3 km in the LMNS data compared to ~0.10 from the NIST reference. Per *Hoffman et al.* (1979b), a $CO^+/CO_2^+$ ratio of ~0.4 was obtained when the gate valve to the ion source in the LMNS was closed. At an altitude of 51.3 km (~ 1 bar), however, we presume that the gate valve was open. Thus, the $CO^+/CO_2^+$ ratio was supportive of $CO^+$ being an atmospheric parent species, where corresponding abundances were disambiguated using the $CO^+/CO_2^+$ ratios from the NIST and LNMS spectra. The data also support the presence of oxygen gas ($O_2$), which *Hoffman et al.* (1980a) attributed to dissociative ionization of $CO_2$. While the NIST spectrum for $CO_2$ (**Figure S3**) shows no formation of $O_2$, the possibility of a ~0.02% yield to form ~320 counts of $O_2^+$ from ~$1.8 \times 10^6$ counts of $CO_2$ could not be excluded. Lastly, using isotopologues of $CO_2$, $N_2$, and SO, and atomic Cl (at select altitudes, **Table S2**), we obtained isotope ratios for $^{13}C/^{12}C$ ($1.33 \times 10^{-2} \pm 0.01 \times 10^{-2}$), $^{15}N/^{14}N$ ($2.63 \times 10^{-3} \pm 0.86 \times 10^{-3}$), $^{18}O/^{16}O$ ($2.18 \times 10^{-3} \pm 0.17 \times 10^{-3}$), $^{33}S/^{32}S$ ($1.4 \times 10^{-2} \pm 0.9 \times 10^{-2}$), $^{34}S/^{32}S$ ($5.8 \times 10^{-2} \pm 0.7 \times 10^{-2}$), and

$^{37}Cl/^{35}Cl$ (4.5x10$^{-1}$ ± 0.7x10$^{-1}$) – which were similar to terrestrial values (**Table S2**) (*Haynes*, 2016; *Farquhar*, 2017).

**Conclusion**

Our assessment of the PV LNMS data supports a composition in the middle clouds that includes the main group hydrides of hydrogen sulfide, phosphine, water, ethane, and possibly ammonia; along with several redox active acids including nitrous acid, nitric acid, sulfuric acid, hydrogen cyanide, and potentially chlorous acid, along with the monoprotic acids of hydrochloric acid and potentially hydrogen fluoride. In total, these assignments illuminate a potential for acid-mediated redox disequilibria within the clouds.

Disequilibria in the lower atmosphere of Venus was discussed by *Florenskii et al.* (1978), in regards to Venera 8 observations of NH$_3$ (*Surkov et al.*, 1973), and by *Zolotov* (1991). These LNMS and Venera 8 observations suggest that disparate chemicals across varying equilibrium states may be sustained by unknown chemistries. We speculate that this includes the injection of appreciable reducing power through volcanism or surface outgassing (*e.g. Ivanov and Head* (2013); *Shalygin et al.* (2015); *Gülcher et al.* (2020)).

Regarding the hypothetical habitability of Venus' clouds, our assignments reveal a potential signature of anaerobic phosphorus metabolism (phosphine), a potential electron donor for anoxygenic photosynthesis (nitrous acid; nitrite) (*Griffin et al.*, 2007), and all major constituents of the terrestrial nitrogen cycle (nitrate, nitrite, possibly ammonia, and N$_2$) (*Madigan et al.*, 2014). Also, the redox pair of nitrate and nitrite support the postulate by *Limaye et al.* (2018) of a hypothetical iron-sulfur cycle in Venus' clouds.

Looking ahead, high-resolution (RP >5000) untargeted and targeted mass spectral approaches through sustained aerial platforms and descending probes would significantly aid in elucidating the gaseous and aerosol compositions in the clouds and atmosphere. The DAVINCI+

mission design concept currently under consideration by NASA serves as an excellent step towards this goal.

**Statement of Contributions**
All authors (RM, SSL, MJW, & JAC) contributed to analysis of the data, assisted in drafting of the report, approved of the submission, and agreed to be accountable for the respective contributions.  RM is the corresponding author.

**Acknowledgments, Samples, and Data**
*Acknowledgements*
It is a pleasure to acknowledge R. Richard Hodges for insightful discussions regarding the LNMS operation, data, and history.  RM acknowledges support from the National Aeronautics and Space Administration (NASA) Research Opportunities in Space and Earth Sciences (NNH18ZDA001N).  SSL was supported by NASA (NNX16AC79G).  MJW was supported by the NASA Astrobiology Program through collaborations arising from his participation in the Nexus for Exoplanet System Science (NExSS) and the NASA Habitable Worlds Program.  MJW also acknowledges support from the GSFC Sellers Exoplanet Environments Collaboration (SEEC), which is funded by the NASA Planetary Science Division's Internal Scientist Funding Model.

*Data Availability*
All LNMS data used in this study were obtained from published reports (*Hoffman et al.*, 1980a; *Hoffman et al.*, 1980b).  The LNMS archive data across all altitudes were posted online on October 8, 2020 by the NASA Space Science Data Coordinated Archive (NSSDC) (https://nssdc.gsfc.nasa.gov/nmc/dataset/display.action?id=PSPA-00649).

**Statement of Conflict of Interest**
All authors declare no conflict of interest.

**Figure Legends**

*Figure 2.* (A) LNMS spectra obtained at 51.3 km with annotations for the major species and in-flight calibrants. (B-F) Approximate peak shapes at 51.3 km obtained from regressions of the mass points at 15 amu ($CH_3^+$), 18 amu ($H_2O^+$), 28 amu ($CO^+$ & $N_2^+$), 40 amu ($^{40}Ar^+$), and 136 amu ($^{136}Xe^+$); y-axis error bars are smaller than marker size of the data points. (G-H) Relationships between calculated amu and $\Delta$amu ($\Delta$amu = calculated amu – expected mass) and full width half maximum (FWHM), where averages and standard deviations (error bars) were calculated across the altitudes between 64.2 and 51.3 km (most error bars are smaller than the marker size); diamonds represent the calculated FWHM from deconvolutions at 31 and 34 amu. (I-J) Fits to the mass pair at 32 amu at 51.3 km showing $^{32}S^+$ (blue), $O_2^+$ (red), and summed value (black) using differing variances for the mass and FWHM terms. (K-N) Fits to the mass pair at 31 amu for $P^+$ (blue), $HNO^+$ (red), and summed value (black) across 59.9 to 51.3 km; x-axis error bars represent the standard deviation for the averaged $\Delta$amu (between 15-40 amu) at the respective altitude, and y-axis error bars represent the square root of the counts. (O-Q) Fits to the mass triplet at 34 amu from 51.3 km for $PH_3$ (red), $H_2S$ (blue), and a composite of $PH_3$ and $H_2S$; plot layout and error bars are as described above. (R-S) Fits to the mass triplet at 34 amu from 50.3 and 55.4 km for a composite of $PH_3$ (red )and $H_2S$ (blue); plot layout and error bars are as described. (T-U) Comparison of fragmentation patterns for $PH_3$ and $H_2S$ from the LNMS data (red circles) and the respective NIST mass spectral references (blue squares); counts for $S^+$, $H_2S^+$, $HDS^+$, $^+P$, $^+PH_3$, and $^+PH_2D$ were obtained as described, while counts for $^+PH_2$ and $HS^+$ were disambiguated using the relative abundances of the parent species; error bars represent the square root of the counts, and masses are displayed in unit resolution for clarity.

**Figure S1.** Gaussian fits to the mass peaks at 15 ($CH_3^+$), 18 ($H_2O^+$), 28 ($CO^+$ & $N_2^+$), 40 ($^{40}Ar^+$), and 136 ($^{136}Xe^+$) amu across the altitudes of 64.2, 61.9, 59.9, 58.3, 55.4, and 51.3 km. Error bars (y-axis) are smaller than the marker size of the data points. For water (18 amu), counts at 17.985 amu were corrected for $^{36}Ar^{++}$ using yields from the NIST mass spectral reference for Ar (and counts for $^{36}Ar^+$); poor fits were obtained at 61.9 and 64.2 km due to relatively higher abundances of $^{36}Ar^{++}$. For 28 amu, contributions from isobaric $CO^+$ and $N_2^+$ were included. For 15, 40, and 136 amu, regressions were minimized using least squares; and for 18 and 28 amu, regressions were minimized by least absolute deviations (LAD). Regressions for $CO^+$ & $N_2^+$ (51.3 km) provided solutions ranging from ~40-60% $CO^+$ and $N_2^+$, where the solution of ~60% $CO^+$ and ~40% $N_2^+$, which is plotted in this Figure, providing the lowest relative summed absolute deviation (SAD); the lower solution of 40% $CO^+$ was used to calculate upper abundances of $NO^+$ in **Section 3.3**.

**Figure S2.** Example fit to the LNMS data at 16 amu for $O^+$ and $CH_4^+$ at 49.4 km; at this altitude, counts are roughly equal, thereby allowing calculation of resolving power between the mass pairs, which was 471 with a 12% valley minima for $O^+$.

**Figure S3**. Comparison of the fragmentation pattern for $CO_2$ from the LNMS data (maroon, positive quadrant) and the NIST mass spectral reference (blue, negative quadrant); corrected $CO^+$ abundances were obtained from simulated spectra, $CO_2^{++}$, $C^{18}O^+$, and $C^{16}O^{18}O^+$ are plotted on the left-hand y-axis, and masses were displayed using unit resolution for clarity.

**Figure S4**. Comparison of fragmentation patterns for $HNO_2$ and $H_2SO_4$ from the LNMS data

(maroon, positive quadrant) and the PubChem and NIST mass spectral references (blue, negative quadrant); masses were displayed using unit resolution for clarity, and relative scales do not reflect the error in the low counts for all potential $H_2SO_4$ fragments.

**Supplemental Methods, Tables, and Results**

*Assignments and Fragmentation Patterns*

Changes to the pre-selected mass values (apparent amu) during operation in the clouds were estimated by tracking pre-selected values that were identical to the exact masses for $CH_3^+$, $H_2O^+$, $CO^+$, and $^{136}Xe^+$, and similar (≤0.003 amu) to the exact masses for $N_2^+$ and $^{40}Ar^+$. Exact masses for $CH_3^+$ and $H_2O^+$ were calculated using a mass of 1.0079 amu for hydrogen, per measurements that were published in 1976 (*Roth et al.*, 1976) before launch of the Pioneer-Venus Large Probe in 1978. Shifts to the pre-selected masses ranged from 0.001-0.007 amu at 51.3 km, 0.000-0.009 amu at 55.4 km, 0.000-0.030 amu at 58.3 and 59.9 km, 0.001-0.023 amu at 61.9 km, and 0.001-0.021 amu at 64.2 km. In practice, we used the maximum shift to assist in sorting initial chemical assignments, where pre-selected masses that differed (absolute) from the expected masses by less than the maximum shift were roughly treated as near-centroids (or near the peak means); while pre-selected values that differed from exact masses by less than the estimated FWHM of the target species were treated as components of the sloping edges of the peak (off-set peak). In turn, regressions to mass pairs and triplets for target species of <40 amu were constrained using the target exact mass and a variance that equaled the averaged $\Delta$amu between 15-40 amu ($CH_3^+$, $H_2O^+$, $CO^+$, $N_2^+$, & $^{40}Ar^+$) at each resepective altitude, along with the estimated FWHM and standard deviation, which was obtained by linear regression at the resepective altitude (similar to **Figure 1H**). Regressions to single mass points, given the uncertainty in the pre-selected mass value, were only used to obtain rough estimates of the calculated maximum counts.

Isobaric species were additionally disambiguated using isotope ratios. Abundances for $^{13}CO^+$ were obtained using the $^{13}C/^{12}C$ ratio ($1.33 \times 10^{-2} \pm 0.01 \times 10^{-2}$) and counts of $CO^+$ from simulated spectra (**Figure 1D** & **S1**); in turn, subtraction of $^{13}CO^+$ from the maximum counts at 29 amu provided abundances for $^{14}N^{15}N$. When considering $N_2^+$ abundances from simulated spectra, this provided a $^{15}N/^{14}N$ ratio of $2.63 \times 10^{-3} \pm 0.86 \times 10^{-3}$ across the altitudes of 61.2-51.3 km ($^{14}N^{15}N$ was below the limit of detection at 64.2 km). For $NO^+$, counts were obtained by (1) using counts for $CO^+$ from simulated plots (**Figure S1** legend), (2) converting to counts for $C^{18}O^+$ using the $^{18}O/^{16}O$ isotope ratio ($2.18 \times 10^{-3} \pm 0.17 \times 10^{-3}$), and (3) constraining regressions to the mass pair at 30 amu (for $NO^+$, $C^{18}O^+$, and $C_2H_6^+$) using the calculated counts of $C^{18}O^+$, estimated FWHM values, and expected masses. Similarly, maximum possible counts for $NO_2^+$ (≤620) were estimated using the error in the $^{18}O/^{16}O$ ratio.

Fragmentation patterns for $CO_2$, $HNO_2$, and $H_2SO_4$ are displayed in **Figures S3-4**. Mass data for parent ions and associated species were binned and plotted against reference spectra obtained from the NIST Chemistry WebBook (https://webbook.nist.gov/chemistry/) (*Wallace*, 2020), MassBank Europe (https://massbank.eu/MassBank/), PubChem (https://pubchem.ncbi.nlm.nih.gov), or from published reports.

*Organic Contamination*

Per *Hoffman et al.* (1980a), pre-flight studies with the LNMS revealed mass signals at 77 and 78 amu that were attributed to benzene arising from the vacuum sealants. Unfortunately, we are aware of no technical reports that describe contamination control for the LNMS. It is also possible that components of the LNMS were cleaned with trichloroethylene (TCE; $C_2HCl_3$) and/or treated with a sealant such as Vacseal® High Vacuum Leak Sealant, which contains TCE, xylene, and ethyl benzene. After assembly, but pre-launch, the LNMS was also possibly subjected to ~750 K for an unknown amount of time to remove organics.

When at Venus, the LNMS performed five complete peak-stepping operations in the upper atmosphere with data collection beginning at 64.2 km. Across the cloud measurements, however, counts and mass values were suggestive of the presence of the TCE parent ion (129.914383 amu), $C_2HCl_3^+$ ($M^+$), along with the ions of $(M+2)^+$, $[M-Cl]^+$, $[(M+2)-Cl]^+$, and possibly $[M-2Cl]^+$. The spread in counts, however, were not consistent with terrestrial $^{37}Cl/^{35}Cl$ ratios, which implied the presence of entangled isobaric species.

Per *Donahue et al.* (1981), these same mass positions (**128.905**, **129.921**, **130.914**, and **131.922 amu**) were assigned to $^{129}Xe$, $^{130}Xe$, $^{131}Xe$, and $^{132}Xe$, which suggested that contamination by TCE was considered to be minimal by the original investigators. Nevertheless, in the event of low-level TCE contamination, then the counts of 5 for the parent ion, $C_2HCl_3^+$, were suggestive of TCE being a minor source of atomic chlorine ($Cl^+$). Per the NIST reference, atomic chlorine is ~10% of the TCE parent ion (base peak), which amounted to hypothetical counts of ~0.5 for atomic chlorine arising from TCE. Reference spectra also indicated that $C_2H_6^+$ and $C_2H_4^+$ – which were potential assignments in the data – were not products of TCE fragmentation.

Evaluation of the NIST reference spectra for *o*-xylene, *m*-xylene, *p*-xylene, and ethyl benzene, indicated that benzene and benzyl radical cations were produced in yields of ~10 and 15% of the base peak (tropylium, $C_7H_7^+$; 91.054775 amu). In the LNMS data, the pre-selected value **78.053 amu** was consistent with the mass of benzene ($C_6H_6^+$), while the counts of 16 implied the presence of a substantially larger base peak and parent ion. However, the LNMS did not sample masses for the xylene and tropylium ions. Again, in the absence of technical information, we are unable to discern between contaminants or the atmosphere as a source of the counts at **78.053 amu**.

We posit that a majority of the residual sealant may have been sufficiently removed by the pre-launch and pre-data acquisition preparations – as may have been the case for TCE. If so, the massive increase at **78.053 amu** to ~30,000 counts at 14-15 km (well below the clouds) may be indicative of alternative chemical species such as dimethyl sulfoxide (DMSO; $(CH_3)_2SO$) or chemical fragments such as $P_2O^+$. In support of this assessment are counts at **78.924 amu**, which likely represent the $^{13}C$-benzene isotopologue, $C_5^{13}CH_6$ (79.050305 amu). In sharp contrast to benzene, counts at this position did not exhibit the massive increase at 14-15 km. Moreover, the relative counts at **78.924**, **78.053**, and **77.040 amu** across the altitude profile were inconsistent with the relative abundances of $C_6H_5^+$, $C_6H_6^+$, and $C_5^{13}CH_6^+$ from the NIST and MassBank references for benzene. Instead, adjustments using the NIST reference provided a maximum of ~870 counts for benzene below the clouds, which was well below the measured counts of ~30,000. At 51.3 km, adjustments of the counts at **78.924** (2), **78.053** (16), and **77.040 (1) amu** were suggestive of maximum values of ~2 counts for $C_6H_5^+$, ~7 counts for $C_6H_6^+$ (benzene), and ~0.5 counts for $C_5{}^{13}CH_6^+$.

**Table S1.** List of parent ions and fragmentation products for (A) $CO_2$, (B) $PH_3$, (C) $H_2S$, (D) $HNO_2$ and $HNO_3$, (E) $H^{35}Cl$ and $H^{37}Cl$, (F) fragments of $H_2SO_4$, (G) $NH_3$, and (H) low-mass organics; where measured counts at the pre-selected masses or calculated (*) counts from simulated spectra are listed.

| (A) carbon dioxide & carbon monoxide ($CO_2$ & $CO$) | | | | |
|---|---|---|---|---|
| apparent amu | count | formula | parent & fragment ions | expected mass |
| 45.995 | 7936 | $CO^{18}O^+$ | $(M+2)^+$ | 45.994160 |
| 44.991 | 21504 | $^{13}CO_2^+$ | $(M+1)^+$ | 44.993355 |
| 43.991 | 1769472 | $CO_2^+$ | $M^+$ | 43.990000 |
| 28.997 | 6656 | $^{13}CO^+$ | CO: $(M+1)^+$ <br> $CO_2$: $[(M+1)-O]^+$ | 28.998355 |
| 29.997 | 801* | $C^{18}O^+$ | CO: $(M+2)^+$ <br> $CO_2$: $[(M+2)-O]^+$ | 29.999160 |
| 27.995 | 423535* | $CO^+$ | CO: $M^+$ <br> $CO_2$: $[M-O]^+$ | 27.995000 |
| 22.496 | 560 | $^{13}CO_2^{++}$ | $(M+1)^{++}$ | 22.496677 |
| 21.995 | 47104 | $CO_2^{++}$ | $M^{++}$ | 21.995000 |
| 15.995 | 335872 | $O^+$ | CO: $[M-C]^+$ <br> $CO_2$: $[M-O-C]^+$ | 15.995000 |
| 12 | 344064 | $^{12}C^+$ | CO: $[M-O]^+$ <br> $CO_2$: $[M-2O]^+$ | 12.000000 |
| (B) phosphine ($PH_3$) | | | | |
| apparent amu | count | formula | parent & fragment ions | expected mass |
| 35.005 | 6* | $^+PH_2D$ | $(M+1)^+$ | 35.003659 |
| 33.992 | 19* | $^+PH_3$ | $M^+$ | 33.997382 |
| 32.985 | 15* | $^+PH_2$ | $[M-H]^+$ | 32.989557 |
| 30.973 | 6* | $P^+$ | $[M-3H]^+$ | 30.973907 |
| (C) hydrogen sulfide ($H_2S$) | | | | |
| apparent amu | Count | formula | parent & fragment ions | expected mass |
| 34.005 | 7* | $HDS^+$ | $(M+1)^+$ | 33.993998 |
| 33.992 | 1.7* | $H_2S^+$ | $M^+$ | 33.987721 |
| 32.985 | 2* | $HS^+$ | $[M-H]^+$ | 32.979896 |
| 31.972 | ≤8* | $^{32}S^+$ | $[M-2H]^+$ | 31.972071 |
| (D) nitrous & nitric acid ($HNO_2$ & $HNO_3$) | | | | |
| apparent amu | count | formula | parent & fragment ions | expected mass |
| 62.994 | 1 | $HNO_3^+$ | $HNO_3$: $M^+$ | 62.995899 |
| 47.000 | 94 | $HNO_2^+$ | $HNO_2$: $M^+$ | 47.000899 |
| 45.995 | ≤620* | $NO_2^+$ | $HNO_3$: $[M-17]^+$ | 45.993074 |

| apparent amu | count | formula | parent & fragment ions | expected mass |
|---|---|---|---|---|
| 31.006 | 26 | HNO$^+$ | HNO$_2$: [M-16]$^+$ | 31.005899 |
| 29.997 | ≤208* | NO$^+$ | HNO$_3$: [M-33]$^+$<br>HNO$_2$: [M-17]$^+$ | 29.998074 |
| 17.002 | 296 | $^+$OH | HNO$_3$: [M-46]$^+$<br>HNO$_2$: [M-30]$^+$ | 17.002825 |
| 15.995 | 335872 | O$^+$ | HNO$_3$: [M-47]$^+$<br>HNO$_2$: [M-31]$^+$ | 15.995000 |
| 14.000 | 19456 | $^{14}$N$^+$ | HNO$_3$: [M-49]$^+$<br>HNO$_2$: [M-33]$^+$ | 14.003074 |

**(E) *hydrochloric acid* (HCl)**

| apparent amu | count | formula | parent & fragment ions | expected mass |
|---|---|---|---|---|
| 37.968 | 36* | H$^{37}$Cl$^+$ | (M+2)$^+$ | 37.973728 |
| 36.966 | 6* | $^{37}$Cl$^+$ | [(M+2)-H]$^+$ | 36.965903 |
| 35.981 | 4* | HCl$^+$ | M$^+$ | 35.976678 |
| 34.972 | 12* | $^{35}$Cl$^+$ | [M-H]$^+$ | 34.968853 |
| 1.008 | 3520 | H$^+$ | [M-Cl]$^+$ & [(M+2)-Cl]$^+$ | 1.007825 |

**(F) *sulfuric acid fragments* (H$_x$SO$_y$; x = 0-2, y = 1-3)**

| apparent amu | count | formula | parent & fragment ions | expected mass |
|---|---|---|---|---|
| 79.958 | 0 | SO$_3^+$ | [M-18]$^+$ | 79.957071 |
| 65.961 | 0.3* | $^{34}$SO$_2^+$ | [(M+2)-34]$^+$ | 64.961459 |
| 64.96 | 3 | HSO$_2^+$ | [M-33]$^+$ | 64.969896 |
| 63.962 | 5 | SO$_2^+$ | [M-34]$^+$ | 63.962071 |
| 50.969 | 0.1* | H$^{34}$SO$^+$ | [(M+2)-49]$^+$ | 50.970692 |
| 49.968 | 3 | $^{34}$SO$^+$ | [(M+2)-50]$^+$ | 49.962867 |
| 48.974 | 2 | $^{33}$SO$^+$ | [[(M+1)-50]$^+$ | 48.966459 |
| 47.966 | 10 | SO$^+$ | [M-50]$^+$ | 47.967071 |

**(G) *ammonia* (NH$_3$)**

| apparent amu | count | formula | parent & fragment ions | expected mass |
|---|---|---|---|---|
| 18.034 | ≤20* | NH$_2$D$^+$ | M$^+$ | 18.032826 |
| 16.018 | 40960 | $^+$NH$_2$ | [M-2H]$^+$ | 16.018724 |
| 15.013 | 7680 | $^+$NH | [M-3H]$^+$ | 15.010899 |
| 14 | 19456 | $^{14}$N$^+$ | [M-4H]$^+$ | 14.003074 |

**(G) *low-mass organics* (C$_x$H$_y$)**

| apparent amu | count | formula | parent & fragment ions | expected mass |
|---|---|---|---|---|
| 78.924 | ~0.5* | C$_5$($^{13}$C)H$_6^+$ | $^{13}$C-Benzene: (M+1)$^+$ | 78.046950 |
| 78.053 | ~7* | C$_6$H$_6^+$ | Benzene: M$^+$ | 78.046950 |
| 77.04 | ~2* | C$_6$H$_5^+$ | Benzene F1: [M-H]$^+$ | 77.039125 |
| 40.029 | 80 | C$_3$H$_4^+$ | Propyne: M$^+$ | 40.031300 |
| 30.046 | ≤100* | $C_2H_6^+$ | Ethane: $M^+$ | 30.046950 |
| 29.039 | 992 | $C_2H_5^+$ | Ethane F1: $[M-H]^+$ | 29.039125 |
| 28.032 | 122880 | $C_2H_4^+$ | Ethene: $M^+$<br>Ethane F2: $[M-2H]^+$ | 28.031300 |
| 27.023 | ≤50* | $C_2H_3^+$ | Ethene F1: $[M-H]^+$<br>Ethane F3: $[M-3H]^+$ | 27.023475 |
| 26.014 | ≤10* | $C_2H_2^+$ | Ethene F2: $[M-2H]^+$<br>Ethane F4: $[M-4H]^+$<br>Ethyne: $M^+$ | 26.015650 |
| 16.031 | 39936 | $CH_4^+$ | Methane: $M^+$ | 16.031300 |
| 15.023 | 22528 | $CH_3^+$ | Methane F1: $[M-H]^+$<br>Ethane F5: $[M-3H-C]^+$ | 15.023475 |

*Calculated or estimated from simulated spectra and/or adjusted using isotope ratios or relative abundances from reference spectra.

**Table S2**

| Isotope Ratios | | | | |
|---|---|---|---|---|
| **Isotopes** | **Venus** | **Altitudes** | **Comments** | **Earth** |
| $^{13}C/^{12}C$ | $1.33 \times 10^{-2} \pm 0.01 \times 10^{-2}$ | 64.2-51.3 km & 23.0-0.9 km | Clog excluded (50.3-24.4 km); $1.28 \times 10^{-2} \pm 0.02 \times 10^{-2}$ was obtained across all altitudes. | $1.08 \times 10^{-2}$ |
| $^{15}N/^{14}N$ | $2.63 \times 10^{-3} \pm 0.86 \times 10^{-3}$ | 59.9-51.3 km | $^{14}N^{15}N_2^+$ was below the detection limit at 64.2 km; and ratio was not calculated <51.3 km. | $3.65 \times 10^{-3}$ |
| $^{18}O/^{16}O$ | $2.18 \times 10^{-3} \pm 0.17 \times 10^{-3}$ | 64.2-51.3 km & 23.0-0.9 km | Clog excluded (50.3-24.4 km); $2.14 \times 10^{-3} \pm 0.26 \times 10^{-3}$ was obtained across all altitudes. | $2.05 \times 10^{-3}$ |
| $^{33}S/^{32}S$ | $1.4 \times 10^{-2} \pm 0.9 \times 10^{-2}$ | 39.3-25.9 km | During the clog where respective counts were enriched. | $7.88 \times 10^{-3}$ |
| $^{34}S/^{32}S$ | $5.8 \times 10^{-2} \pm 0.7 \times 10^{-2}$ | 39.3-25.9 km | During the clog where respective counts were enriched. | $4.39 \times 10^{-2}$ |
| $^{37}Cl/^{35}Cl$ | $4.5 \times 10^{-1} \pm 0.7 \times 10^{-1}$ | 58.3-51.3 km | $^{37}Cl^+$ was below the detection limit >58.3 km; and ratio not calculated <51.3 km. | $3.20 \times 10^{-1}$ |

**Figure S1**

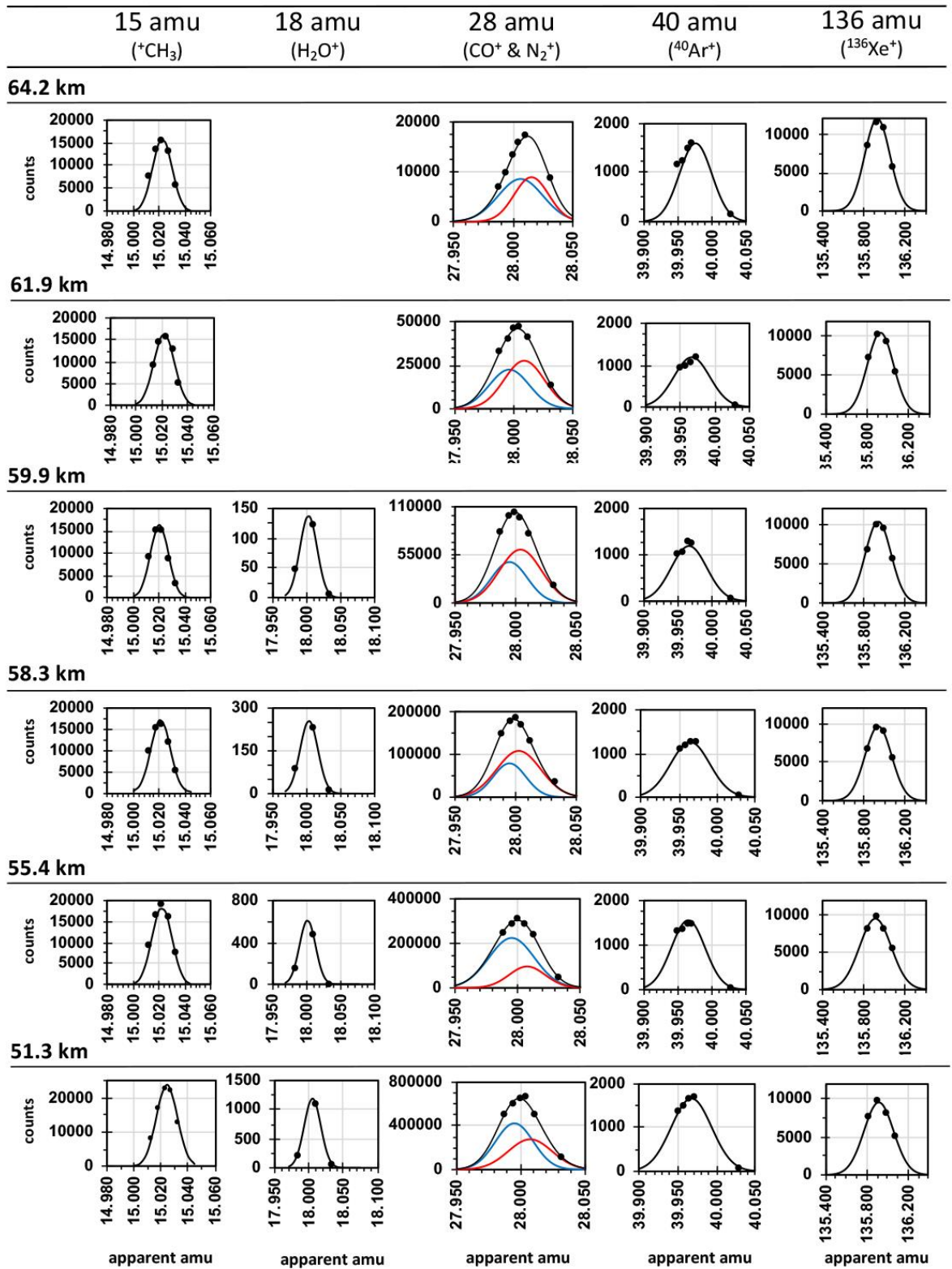

**Figure S1.** Gaussian fits to the mass peaks at 15 ($CH_3^+$), 18 ($H_2O^+$), 28 ($CO^+$ & $N_2^+$), 40 ($^{40}Ar^+$), and 136 ($^{136}Xe^+$) amu across the altitudes of 64.2, 61.9, 59.9, 58.3, 55.4, and 51.3 km. Error bars (y-axis) are smaller than the marker size of the data points. For water (18 amu), counts at 17.985 amu were corrected for $^{36}Ar^{++}$ using yields from the NIST mass spectral reference for Ar (and counts for $^{36}Ar^+$); poor fits were obtained at 61.9 and 64.2 km due to relatively higher abundances of $^{36}Ar^{++}$. For 28 amu, contributions from isobaric $CO^+$ and $N_2^+$ were included. For 15, 40, and 136 amu, regressions were minimized using least squares; and for 18 and 28 amu, regressions were minimized by least absolute deviations (LAD). Regressions for $CO^+$ & $N_2^+$ (51.3 km) provided solutions ranging from ~40-60% $CO^+$ and $N_2^+$, where the solution of ~60% $CO^+$ and ~40% $N_2^+$, which is plotted in this Figure, providing the lowest relative summed absolute deviation (SAD); the lower solution of 40% $CO^+$ was used to calculate upper abundances of $NO^+$ in **Section 3.3**.

**Figure S2**

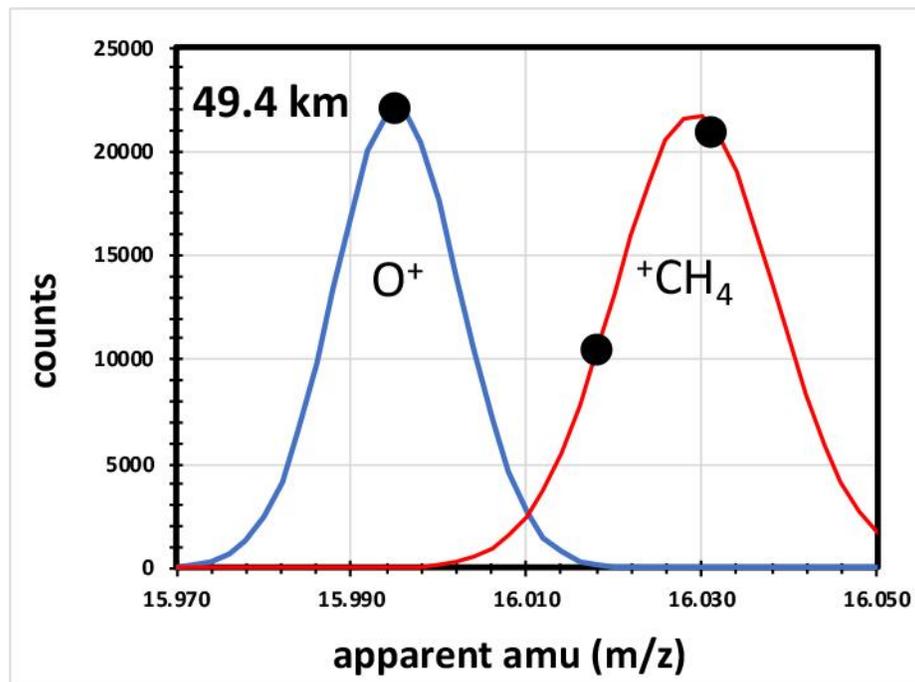

**Figure S2.** Example fit to the LNMS data at 16 amu for O⁺ and CH₄⁺ at 49.4 km; at this altitude, counts are roughly equal, thereby allowing calculation of resolving power between the mass pairs, which was 471 with a 12% valley minima for O⁺.

**Figure S3**

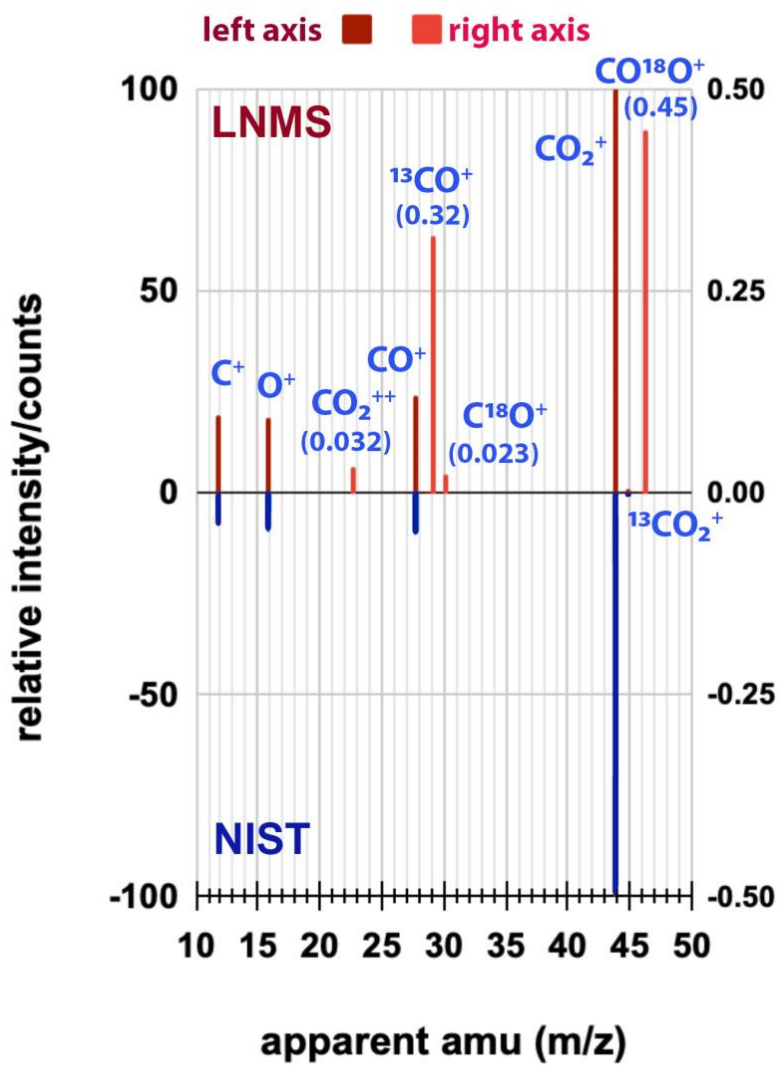

**Figure S3.** Comparison of the fragmentation pattern for $CO_2$ from the LNMS data (maroon, positive quadrant) and the NIST mass spectral reference (blue, negative quadrant); corrected $CO^+$ abundances were obtained from simulated spectra, $CO_2^{++}$, $C^{18}O^+$, and $C^{16}O^{18}O^+$ are plotted on the left-hand y-axis, and masses were displayed using unit resolution for clarity.

**Figure S4**

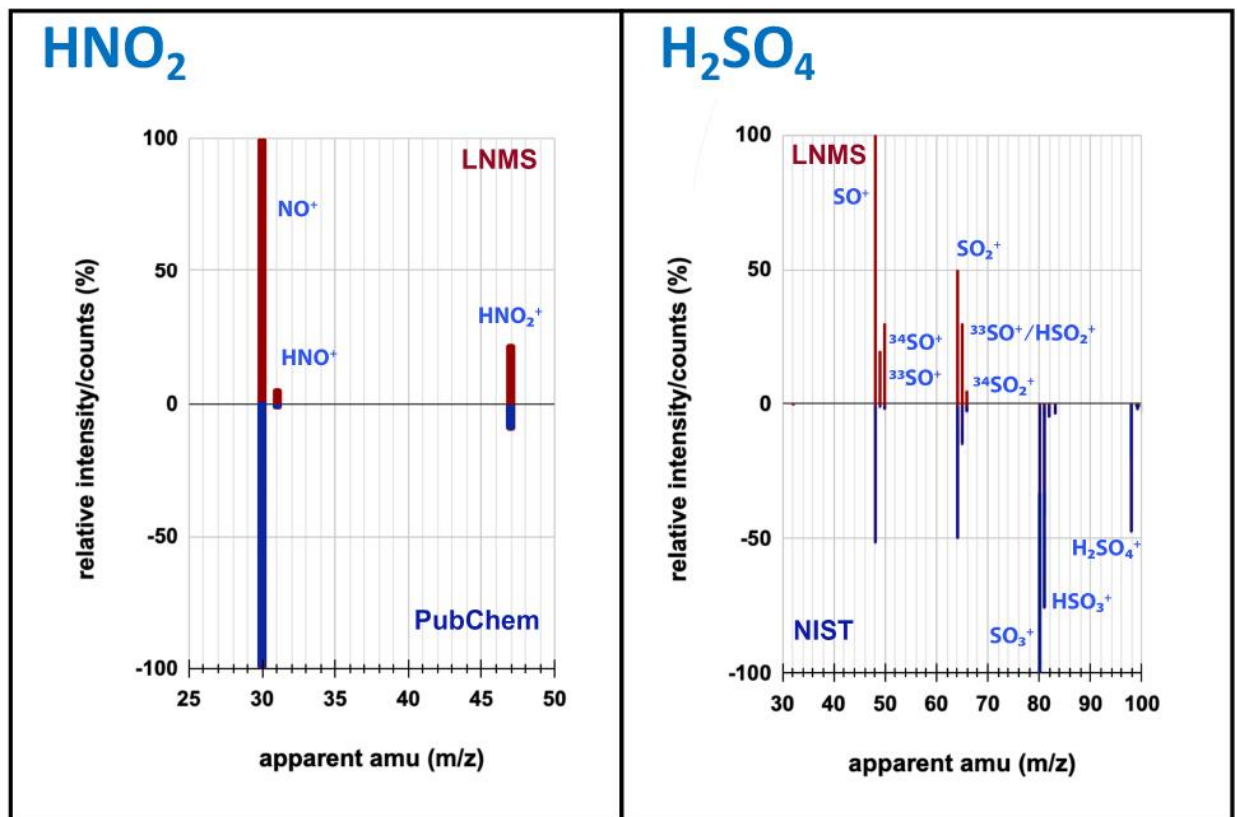

**Figure S4**. Comparison of fragmentation patterns for HNO$_2$ and H$_2$SO$_4$ from the LNMS data (maroon, positive quadrant) and the PubChem and NIST mass spectral references (blue, negative quadrant); masses were displayed using unit resolution for clarity, and relative scales do not reflect the error in the low counts for all potential H$_2$SO$_4$ fragments.